\documentclass[twocolumn,showpacs,preprintnumbers,amsmath,amssymb]{revtex4}

\usepackage{graphicx}
\usepackage{dcolumn}
\usepackage{bm}

\begin{document}
\title{The synchronizability of highly clustered scale-free networks}
\author{Xiang Wu}
\author{Bing-Hong Wang}
\email{bhwang@ustc.edu.cn}
\author{Tao Zhou}
\author{Wen-Xu Wang}
\author{Ming Zhao}
\author{Hui-Jie Yang}
\affiliation{%
Department of Modern Physics and Nonlinear Science Center,
University of Science and Technology of China, Hefei, 230026, PR
China
}%

\date{\today}

\begin{abstract}
In this letter, we consider the effect of clustering coefficient on
the synchronizability of coupled oscillators located on scale-free
networks. The analytic result for the value of clustering
coefficient aiming at a highly clustered scale-free network model,
the Holme-Kim is obtained, and the relationship between network
synchronizability and clustering coefficient is reported. The
simulation results strongly suggest that the more clustered the
network, the poorer its synchronizability.

\end{abstract}


\maketitle Many social, biological, and communication systems can be
properly described as complex networks with nodes representing
individuals and edges mimicking the interactions among them
\cite{Review1,Review2,Review3}. Examples are numerous: these include
the Internet, the World Wide Web, social networks, metabolic
networks, food webs, and many others
\cite{Li2005,Zhang2005,CPL1,CPL2}. Recent empirical studies indicate
that the networks in various fields have some common
characteristics, the most important of which are called small-world
effect \cite{WS} and scale-free property \cite{BA}. Networks of
small-world effect have small average distance as random networks
and large clustering coefficient as regular ones. And the scale-free
property means the degree distribution of networks obeys the
power-law form.

One of the ultimate goals of researches on complex networks is to
understand how the structure of complex networks affects the
dynamical process taking place on them, such as traffic flow
\cite{Traffic1,Traffic2,Traffic3,Traffic4}, epidemic spread
\cite{Epidemic1,Epidemic2,Epidemic3}, cascading behavior
\cite{Cascade1,Cascade2,Cascade3}, and so on. In this letter, we
concentrate on the synchronization, which is observed in a variety
of natural, social, physical and biological systems
\cite{SYNC1,SYNC2,SYNC3}. The large networks of coupled dynamical
systems that exhibit synchronized state are subjects of great
interest. Previous studies mainly focus on the Wastts-Strogatz
\cite{WS} networks and Barab\'{a}si-Albert \cite{BA} networks, which
have demonstrated that scale-free and small-world networks are much
easier to synchronize than regular lattices
\cite{Easier1,Easier2,Easier3,Easier4}. Since many real-life
networks are scale-free small-world networks, and there are already
some models that can simultaneously reproduce the small-world and
scale-free characteristics \cite{SS1,SS2,SS3,SS4}, to investigate
the synchronizability of the scale-free small-world networks if of
great interest and importance.

Since the scale-free networks are always of very small average
distances \cite{Cohen2003}, the scale-fee small-world networks can
also be referred as highly clustered scale-free networks. One of
the earliest highly clustered scale-free models is the Holme-Kim
(HK) model \cite{SS1}, which has successfully reproduced the
indirectly acquainting mechanism in real networks thus is closer
to reality than BA model. In this letter, we will firstly give the
analytical result about the clustering coefficient of HK model,
which is helpful for understanding the underlying evolution
mechanism of HK model. And then, we will investigate the
relationship between network synchronizability and clustering
coefficient based on HK model.

As a remark, previous studies mainly concentrate on how the average
distance and heterogeneity of degree/betweenness distribution affect
the network synchronizability
\cite{Easier2,Easier3,Zhou2005,Zhao2005,Betweenness1,Betweenness2},
while there are few systemic works about the effect of clustering
coefficient. Although there are a number of highly scale-free
models, the HK model is a typical one who has tunable clustering
coefficient thus provides us a good researching stage. This is the
reason why we choose HK model as our theoretic template.

The HK network is generated by the following processes:

(1) In each step, $m$ edges are added in the networks, and $t$ is a
discrete parameter which denotes the global time that the system
totally goes.

(2) First an edge is add with the probability $\Pi(k_{i})$,
\begin{equation}
\Pi(k_{i})=\frac{k_{i}}{\sum_{j}^{ }k_{j}}.
\end{equation}

(3) Then, in the following $m-1$ time steps, do a PA (preferential
attachment) step with the probability $p$ or a TF (triad formation)
step with the probability $1-p$ (see Ref.[27] for details).

By using the rate-equation \cite{13}, one can obtain the evolution
of nodes' degree as follows:

\begin{eqnarray}
\frac{\partial k_i(t)}{\partial
t}&=&\{1-(1-\frac{k_i(t-1)}{\sum_{j=1}^{t-1}k_j(t-1)})^{m_{PA}}\}
\nonumber\\
&+&\{1-(1-\frac{\sum_{l\in\Omega_i}k_l(t-1)\frac{1}{k_l(t-1)}}{\sum_{j=1}^{t-1}k_j(t-1)})^{m_{TF}}\}.
\end{eqnarray}
where $\Omega_i$ denotes the set of neighbors of  node $i$,
$k_{i}(t)$ is the degree of node $i$ at time step $t$. In the above
formula, $m_{TF}$ is the number of edges that is connected following
the rule of triad formation in each step while $m_{PA}$ denotes the
number of those connected following the rule of preferential
attachment. Denote $k{_i}:=k{_i}(N)$, where $N$ is the network size,
the two terms in the right side of the above formula are the degree
increment rate of node $i$ in PA and TF step, respectively. By using
the initial condition $k_{i}(t_{i})=m$, and the expressions of
$m_{PA}=(m-1)(1-p)+1$ and $m_{TF}=p(m-1)$, one can obtain the
solution as follows:

\begin{equation}
k_{i}(t)=k_{i}(t;t_{i},m),
\end{equation}
from which and by using of the continuum theory \cite{12}, one can
get

\begin{equation}
P(k,p)=\frac{2m^2}{k^3}+A\frac{m^2}{k^{2}N},
\end{equation}
where $A$ is a quadric polynomial of $p$. Clearly, $p(k)=k^{-3}$ in
the limit case $N\rightarrow\infty$.

The clustering coefficient of the whole network is the average of
$c_i$ over all nodes $i$, where $c_i$ is the ratio between the
number of edges among node $i$'s neighbors which is denoted by
$n_i$ and the total possible number. Then,

\begin{equation}
c_{i}=\frac{2n_i}{k_i(k_i-1)}.
\end{equation}

Using the rate-equation approach \cite{13}, the detailed expression
of $c(k)$, which denotes the average clustering coefficient over all
the $k$-degree nodes, should be as follows:

\begin{eqnarray}
c(k)&=&\frac{1}{k(k+1)}\{\frac{2m_{PA}m_{TF}}{m}(k-m) \nonumber\\
&+&\frac{1}{k(k+1)}\{\frac{2m_{PA}m_{TF}}{m}(k-m) \nonumber\\
&+&\frac{m_{PA}(m_{PA}-1)}{16m} \frac{(\ln N)^2}{N}k^2 \}.
\end{eqnarray}
Here we assume that $m_{0}$ (the number of initial nodes) is equal
to $m$ (the edges added each time step). The three items in the
right side in the above expression is got from three mechanism
shown in Fig. 1.

\begin{figure}
\scalebox{0.50}[0.50]{\includegraphics{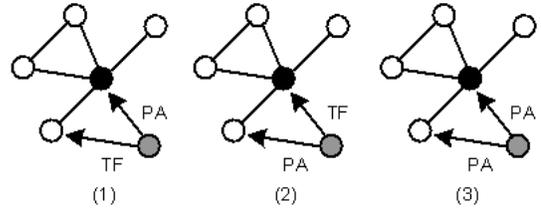}}
\caption{\label{fig:epsart}The three parts of this figure represent
three mechanisms as: (1) node $i$ is connected by a PA step while
the neighbors of it is also connected by a TF step; (2) one of node
$i$'s neighbors is connected by a PA step while the node $i$ is
connected by a TF step; (3) node $i$ as well as one of its neighbors
is connected by a PA step respectively.}
\end{figure}

Thus the clustering coefficient $C$ can be solved as a function of
the free parameter $p$,

\begin{equation}
c(p)=\int_{k_{min}}^{k_{max}}c(k)P(k)\mathrm{d}k.
\end{equation}
In the above formula, $k_{max}\rightarrow2m\sqrt{N}$ and $k_{min}=m$
\cite{Cohen2000}. For $p\in[0,1]$, $c(p)$ can be simplified in
linear approximation.

\begin{equation}
c(p)=B(m,N)+c(m,N)p.
\end{equation}
The extensive simulation results with different $p$ and $c$ for
networks of different sizes strongly support the analytic results,
especially in the larger-size networks, as shown in Fig. 2.

\begin{figure}
\scalebox{0.80}[0.80]{\includegraphics{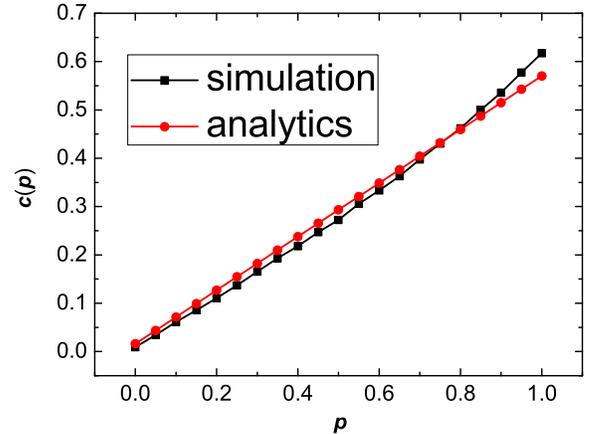}}
\caption{\label{fig:epsart}Clustering coefficient $c(p)$ vs the
parameter $p$ in HK model. The square-dot curve represents the
simulation result while the circle-dot curve denotes the analytic
solution. The network size is $N=5000$, and $m=m_{0}=10$.}
\end{figure}

Other simulations have been done about the variation of average
path length $l$ and standard deviation of degree distribution
$\sigma$. The results show that both the average path length $l$
and standard deviation of degree distribution $\sigma$ behave
slightly variation with vary $p$. The intuitionistic explanation
is quite easily understood that $\sigma$ is directly related with
degree distribution, that is to say, it is determined totally by
the degree distribution, and hardly will the degree distribution
vary when is $N$ large enough. So we have the reason to neglect
the effect of the two structural properties on synchronizability.

Here, we concern the system of linear coupled limit-cycle
oscillators on HK networks. Describing the state of the $i$th
oscillator by $x{_i}$, the equations of motion governing the
dynamics of the $N$ coupled oscillators are:

\begin{equation}
\dot{x}{_i}=F(x{_i})+K\sum_{j=1}^{N}M_{ij}G(x_j),
\end{equation}
where $\dot{x}{_i}=F(x{_i})$ characterizes the dynamics of
individual oscillators, $G(x_j)$ is the output function, $K$
denotes the coupling strength and the $N\times N$ coupling matrix
$M$ is

\begin{equation}
M_{ij}=\left\{
    \begin{array}{cc}
        -k_i, &\mbox{for $i=j$}\\
        1, &\mbox{for $j\in \Lambda_i$}\\
        0, &\mbox{otherwise.}
    \end{array}
    \right.
\end{equation}
In the above expression, $\Lambda_i$ denotes the neighbors of node
$i$. Because of the negative semidefinition and the zero sum of
each raw of the matrix, all its eigenvalues are nonpositive real
values and the biggest eigenvalue $\lambda_0$ is always zero. Thus
the eigenvalues can be ranked as $\lambda_0 \geq \lambda_1 \geq
\ldots \geq \lambda_N$, and $\lambda_0 = \lambda_1 = 0$ if and
only if the network is disconnected.

In our coupled dynamic network, all the oscillators are identical
and the same output function is used, the coupling fashion ensures
the synchronization manifold is an invariant manifold and the
nodes can be well approximated near the synchronous state by a
linear operator. Under these conditions, the eigenratio
$R=\frac{\lambda_{N}}{\lambda_{2}}$ can be used to measure the
network synchronizability; the smaller it is, the stronger the
synchronizability\cite{Easier3,Zhao2005,Betweenness1,Betweenness2,ZhaoZhouWang,syn1,syn2,syn3,syn4,syn5,syn6,syn7}.

We take only synchronizability $R(p)$ and clustering coefficient
$c(p)$ into consideration. Having simulated $R(p)$ for different
configurations versus $p$ (see Figure 3), we know that $R(p)$ is
positively correlated with $p$, and so is $c(p)$ although it is
not completely linear with $p$. Besides, we can easily find that
when $p$ is not very large, the curve is approximately linear (see
Figure 4).

\begin{figure}
\scalebox{0.80}[0.80]{\includegraphics{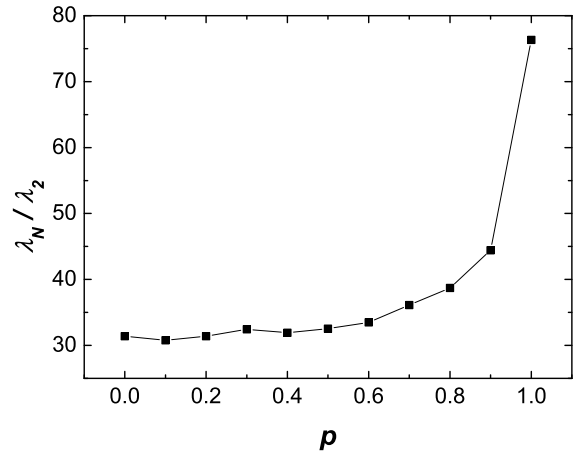}}
\caption{\label{fig:epsart} Synchronizability $R$ measured by
$\frac{\lambda_{N}}{\lambda_{2}}$ vs the parameter $p$ in HK model.
The simulation has been done under the condition that the size of
the networks is $n=1800$. Other parameters are $m=m_{0}=10$.}
\end{figure}

\begin{figure}
\scalebox{0.80}[0.80]{\includegraphics{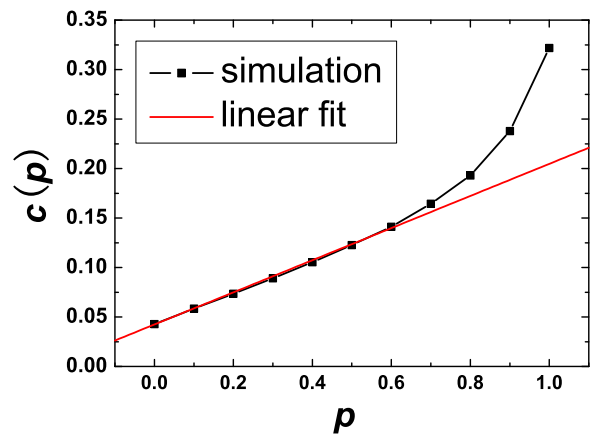}}
\caption{\label{fig:epsart} Clustering coefficient $c(p)$ vs
parameter $p$ under the condition of $N=1800$, and $m=m_0=10$. The
linear fitted line is for the first seven points.}
\end{figure}

Furthermore, we report the relationship between synchronizability
$R(p)$ and clustering coefficient $c(p)$ as shown in Fig. 5. From
the above figure, one important fact the curve reveals is
exponential growing tendency. The function of fitted curve can be
set as follows:

\begin{figure}
\scalebox{0.80}[0.80]{\includegraphics{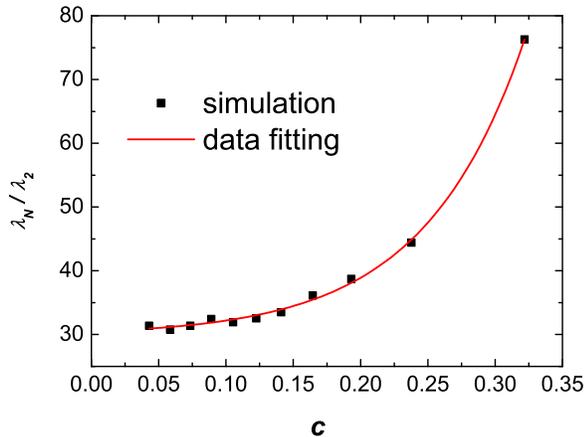}}
\caption{\label{fig:epsart} Synchronizability $R$ measured by
$\frac{\lambda_{N}}{\lambda_{2}}$ as a function of clustering
coefficient $c$ in the case of $N=1800$, and $m=m_0=10$. The fitted
curve follows the tendency of exponential growth.}
\end{figure}

\begin{equation}
R(c)=A(m,N)+B(m,N)e^{\frac{c}{T(m,N)}}.
\end{equation}
The terms $A(m,N),B(m,N)$, and $T(m,N)$ can be obtained by
simulation. Here the simulation averaged over 50 different
realizations is to measure the effect of random fluctuation of
degree distribution, since the degree distribution can both affect
$l$ and $\sigma$. In our simulation, due to the average effect,
the two parameters, $l$ and $\sigma$ do not vary significantly so
that we can only focus on clustering coefficient exclusively which
causes the change of synchronizability.

Compared to the previous works, the advantages of HK model is that
the clustering coefficient can be tuned while the other structural
properties are almost kept fixed. Therefore, combining the
behaviors of $c(p)$ versus $p$ and $R(p)$ versus $p$, we obtain
the relationship between synchronizability and clustering
coefficient. Fig. 4 demonstrates that the larger the clustering
coefficient, the poorer the synchronizability. Due to that the
synchronizability $R$ is determined by the ratio of maximal and
minimal eigenvalues of coupling matrix which is exclusively
related to the network topology, moreover $c$ is the only varied
topological property, thus clustering coefficient plays a crucial
role in synchronizability. The size effect has not been discussed
so far in this letter. As the size $N$ increases the average
distance will get larger, as a result, the synchronization will
become harder, but there will presumably not occur any qualitative
changes, especially on the negative correlation between clustering
and synchronizability. Preliminary simulation results strengthen
this conjecture.

To ascertain the effect of each structural characteristics on
synchronizability is a meaningful work because if future study can
ascertain the relationships between each typical structural
characteristics ($l$, $\sigma$ etc.) and synchronizability
exclusively, we can finally get the expression of
synchronizability as function of those properties. Thus whether a
network can achieve synchronization, and how is synchronizability
of a specific structure can be easily predicted only by the
topological characteristics of it.

This work is support by the National Natural Science Foundation of
China under Nos. 70271070, 10472116, 70471033 and 70571074, and the
Specialized Research Fund for the Doctoral Program of Higher
Education under No. 20020358009.

\end{document}